\documentstyle[12pt]{article}

\newcommand{\be}{\begin{equation}}
\newcommand{\ee}{\end{equation}}
\newcommand{\bea}{\begin{eqnarray}}
\newcommand{\eea}{\end{eqnarray}}

\def\double{\baselineskip 18pt \lineskip 10pt}

\begin{document}

\vspace{1in}

\begin{center}
\Large
{\bf Black holes and gravitational waves in concert --- a probe of 
superstring cosmology}

\vspace{0.5in}

\normalsize

\large{Edmund J.~Copeland$^1$, Andrew R.~Liddle$^2$, James
E.~Lidsey$^{1,2}$ and David Wands$^3$}

\normalsize
\vspace{.4in}

{\em $^1$Centre for Theoretical Physics, University of Sussex,\\
Brighton BN1 9QJ, Great Britain}

\vspace*{12pt}

{\em $^2$Astronomy Centre, University of Sussex,\\ 
Brighton BN1 9QJ, Great Britain}

\vspace*{12pt}

{\em $^3$School of Computer Science and Mathematics, University of 
Portsmouth,\\ Portsmouth PO1 2EG, Great Britain}

\end{center}

\begin{abstract}
Two strands of observational gravitation, one the search for astrophysical
evidence of primordial black holes and the other the
search for gravitational waves, may combine to provide strong evidence in
favour of cosmological models based on superstring theory, the leading 
candidate for unifying gravity with the other fundamental forces.
\end{abstract}


\begin{center}
{\em Awarded Fifth Prize in the 1998 Gravity Research Foundation Essay
Competition.\\
To appear in General Relativity and Gravitation}
\end{center}

\double 

The quest for a unified theory of the fundamental interactions, including
gravity, is the outstanding goal of modern physics.  Superstring theory is
currently the favoured candidate for such a theory, and as such it should
describe the evolution of the very early universe.  The primordial spectrum
of perturbations generated during that period, provides a means for
observationally constraining such a theory, on energy scales that are
inaccessible to any form of terrestrial experiment.  In this essay, we show
that the astrophysical effects of evaporating primordial black holes,
together with a stochastic background of primordial gravitational waves,
will, if observed, provide strong support for inflationary models within
superstring theory \cite{CLLW}.

String theory has undergone a revolution in recent years (see
\cite{GIBBONS}  for an  entertaining review of the most recent
developments).  It is now widely believed that the five separate
perturbative theories are related non--perturbatively by discrete
`duality' symmetries. One such duality is T--duality, relating a theory
compactified on a space of large volume with one compactified on a space
of small volume. The application of T--duality to cosmology has recently
led to a new inflationary scenario, the so--called {\em pre--big bang}
string cosmology \cite{pbb}. 

Inflation is a central paradigm of early universe cosmology \cite{KT}.  It
postulates the existence of a finite, but very rapid, period of accelerated
expansion in the universe's distant past.  Although inflation was originally
developed to explain a number of puzzles of the standard hot big bang model,
by far its most important feature is the generation of scalar (density) and
tensor (gravitational wave) perturbations from quantum vacuum fluctuations.
Small-scale fluctuations generated during inflation are stretched beyond the
Hubble radius by the cosmic expansion, where their amplitude remains frozen
until they re-enter during the radiation or matter dominated epochs.

A much-advertised prediction of string cosmology is that the spectrum of
gravitational waves could be observed by the next generation of
gravitational wave detectors, such as the Laser Interferometric
Gravitational Wave Observatory (LIGO) currently under
construction~\cite{GG,AB}. In a pre--big bang phase driven by the
dilaton field of string theory, the spacetime curvature grows rapidly.
As a consequence the spectrum of gravitational wave perturbations grows 
rapidly
towards higher frequencies, scaling as $f^3$ where $f$ is frequency. The 
current frequency of these gravitational waves depends on the cosmological 
model, but reasonable assumptions place the highest frequency $f_{{\rm s}}$, 
corresponding to the horizon scale at the end of the dilaton phase, around 
the frequencies accessible to LIGO \cite{AB}.
The potentially high amplitude of these waves is in contrast to conventional 
models of inflation, where the
gravitational wave spectrum must slowly decrease  with increasing
frequency. In this latter class of models, the microwave background
anisotropies detected on large angular scales by the Cosmic Background
Explorer (COBE) satellite then leads to an upper limit on the amplitude
of the perturbations that is many orders of magnitude below the maximum
sensitivity of even advanced versions of the LIGO
configuration~\cite{ARL}. The possibility of detecting
inflation-generated gravitational waves is therefore a characteristic
and distinctive feature of the pre--big bang scenario.

Another striking feature of the pre--big bang scenario is that the
gravitational wave amplitude can be related to the probability of black
hole formation on a given scale. We define the scalar and tensor
perturbations, $A_{\rm S}^2$ and $A_{\rm T}^2$, following the
conventions of Ref.~\cite{LLKCBA}. The perturbations produced during the
dilaton phase of the pre--big bang cosmology are related by the
exact equation $A_{\rm T}^2=3A_{\rm S}^2$ \cite{CLLW}. The energy
density of gravitational waves at the present epoch is given in terms
of the original amplitude from the expression 
\be 
\label{Omega} 
\Omega_{\rm gw} (k) = \frac{25}{6} \frac{A^2_{\rm T} (k)}{z_{\rm eq}}
\ee 
where $z_{\rm eq} = 24\,000\,\Omega_0 h^2$ is the redshift of
matter--radiation equality, $\Omega_0$ and $h$ are the present
density and Hubble parameters in the usual units, and $k$ is the wavenumber 
(interchangeable with frequency $f$ as we set $c=1$ throughout). This implies 
that 
\be
\label{ratioomega} 
A^2_{\rm S} =\frac{1}{3} A^2_{\rm T} =2\times 10^{-3} \, 
	\frac{\Omega_{\rm gw}}{10^{-6}} \,\Omega_0 h^2 
\ee 
The advanced LIGO configuration will be sensitive to $\Omega_{\rm gw}
\approx 10^{-9}$ over a range of scales around $100\,{\rm Hz}$.

Density perturbations on very small scales are constrained, because large
inhomogeneities lead to the formation of tiny primordial black holes through
immediate gravitational collapse of the perturbations once they enter the
horizon.  The subsequent Hawking evaporation of these objects results in
numerous astrophysical effects.  That these effects have yet to be observed
places limits on the original number density of black holes that form and, by
implication, the amplitude of the original perturbations.  Larger black holes
will not have evaporated by the present day, but their number density is
constrained by their contribution to the overall matter density in the
universe.

The criterion for a region to collapse into a black hole during the
radiation--dominated epoch is that the density contrast at reentry
should exceed some critical value, around $\delta_{\rm c} = 1/3$. The
mass of the black hole which forms is comparable to the horizon mass at
that time. If a comoving scale $f_*$ reenters the Hubble radius when the
temperature is $T_*$, one can show that 
\be
\label{show}
\frac{f_*}{f_0} \approx \frac{T_*}{T_{\rm eq}}z^{1/2}_{\rm eq}
\ee
where $f_0 \approx H_0 \approx 10^{-18}\,{\rm Hz}$ is the minimum 
observable frequency, corresponding to one oscillation in the lifetime
of the present Universe. Since $T_{\rm eq} \approx 10^4T_0 \approx
1 {\rm eV}$, it follows that
\be
\label{follows}
\frac{f_*}{100\,{\rm Hz}} \approx \frac{T_*}{10^9\, {\rm GeV}}
\ee
The horizon mass at a given temperature is $M_{\rm hor} \approx 
10^{32} \,(T/{\rm GeV})^{-2}\,{\rm g}$, yielding a black hole mass for
a given mode $f_*$ of 
\be
\label{bhmass}
M\approx 10^{14} \left( \frac{100\,{\rm Hz}}{f_*} 
 \right)^2 {\rm g}
\ee
Primordial black holes with initial masses of the order $10^{14}{\rm g}$
are at the final stages of their evaporation today. It is intriguing
that this mass scale corresponds to frequencies observable by LIGO. 

The probability of black hole formation is determined by the 
dispersion, $\sigma_{\rm hor}$, of the matter distribution smoothed 
over the length scale $R \approx f^{-1}$ when that scale re-enters the
Hubble radius. This dispersion can be obtained directly from the power
spectrum $A_{{\rm S}}^2$ \cite{CLLW}, and using Eq.~(\ref{ratioomega}) is
related to the gravitational wave amplitude by
\be
\label{dispersion}
\sigma^2_{\rm hor} = 3 \times 10^{-3} \, \Omega_0 h^2 
 \frac{\Omega_{\rm gw}^{\rm s}}{10^{-6}}
\ee
where $\Omega^{\rm s}_{\rm gw} \equiv \Omega_{\rm gw} (f_{\rm s})$. The
fraction of the mass of the universe, $\beta$, collapsing into black
holes is obtained from the volume of the Universe where the threshold
$\delta_{{\rm c}}$ is exceeded. This is given by the fraction of the
gaussian distribution above $\delta_{\rm c}$, which corresponds to $\beta
={\rm erfc} (\delta_{\rm c}/\sqrt{2}\,\sigma_{\rm hor})$, where `erfc' is
the complementary error function.

Black holes with mass greater than $10^9{\rm g}$ evaporate after one
second and the observational constraints are well known \cite{cgl}. 
Only a
tiny fraction of the mass of the universe may form primordial black
holes; a robust upper limit on the allowed initial mass fraction is
$\beta (M) < 10^{-20}$ and this implies that $\sigma_{\rm hor} < 0.04$.
Saturating the observational bound on $\sigma_{\rm hor}$ gives, from 
Eq.~(\ref{dispersion}), the amplitude of gravitational waves at frequency 
$f_{{\rm s}}$ that
would lead to astrophysical effects from primordial black holes: 
\be
\label{observable}
\Omega_{\rm gw}^{{\rm s}} =\frac{5 \times 10^{-6}}{\Omega_0 h^2}
\ee
This is below the present tightest bound on the gravitational wave
background, which comes from its effect on nucleosynthesis and requires
$\Omega_{\rm gw} < 5 \times 10^{-5}$. But it is well within the
sensitivity of the advanced LIGO detectors~\cite{AB}. 

We stress that a number of assumptions
go into this result, discussed in detail in Ref.~\cite{CLLW}. The peak
of the perturbation spectra arising from the pre--big bang scenario  is
assumed to lie in the frequency range accessible to LIGO. In general
this need not be so, and could lead to a much reduced level of both
gravitational waves and density perturbations on these scales. The crucial 
point though is that the amplitudes are so closely
linked, and if one of the two spectra is observable there is reasonable
hope that the other will be too. For example, if gravitational waves are
detected at around the level of Eq.~(\ref{observable}), then string
cosmology predicts that black holes be observable. Detection of a
gravitational wave background above the level indicated in
Eq.~(\ref{observable}), without black hole detection, would suggest that
these gravitational waves could not have been generated by the dilaton
phase of string cosmology. Detection of the two in concert, with the
correct relation between their amplitudes, would provide possibly the
first observational evidence for string theory.

\end{document}